# Impact of evidence-based flipped or active-engagement non-flipped courses on student performance in introductory physics

Nafis I. Karim[1], Alexandru Maries[2] and Chandralekha Singh[1]
[1] *Department of Physics and Astronomy, University of Pittsburgh, Pittsburgh, PA 15260*
[2] *Department of Physics, University of Cincinnati, Cincinnati, OH 45221*

**Abstract**. We describe the impact of physics education research-based pedagogical techniques in flipped and active-engagement non-flipped courses on student performance on validated conceptual surveys. We compare student performance in courses which make significant use of evidence-based active engagement (EBAE) strategies with courses that primarily use lecture-based (LB) instruction. All courses had large enrollment and often had 100-200 students. The analysis of data for validated conceptual surveys presented here includes data from large numbers of students from two-semester sequences of introductory algebra-based and calculus-based introductory physics courses. The conceptual surveys used to assess student learning in the first and second semester courses were the Force Concept Inventory and the Conceptual Survey of Electricity and Magnetism, respectively. In the research discussed here, the performance of students in EBAE courses at a particular level is compared with LB courses in two situations: (I) the same instructor taught two courses, one of which was a flipped course involving EBAE methods and the other an LB course, while the homework, recitations and final exams were kept the same, (II) student performance in all of the EBAE courses taught by different instructors were averaged and compared with LB courses of the same type also averaged over different instructors. In all cases, we find that students in courses which make significant use of active-engagement strategies, on average, outperformed students in courses using primarily lecture-based instruction of the same type on conceptual surveys even though there was no statistically significant difference on the pretest before instruction. We also discuss correlation between the performance on the validated conceptual surveys and the final exam, which typically placed a heavy weight on quantitative problem solving.

## I. INTRODUCTION

### A. Physics Education Research-based Active Engagement Methods:

In the past two decades, physics education research has identified the challenges that students encounter in learning physics at all levels of instruction [1-15]. Building on these investigations, researchers are developing, implementing and evaluating evidence-based curricula and pedagogies to reduce these challenges to help students develop a coherent understanding of physics concepts and enhance their problem solving, reasoning and meta-cognitive skills [16-27]. In evidence-based curricula and pedagogies, the learning goals and objectives, instructional design, and assessment of learning are aligned with each other and there is focus on evaluating whether the pedagogical approaches employed have been successful in meeting the goals and enhancing student learning.

One highly successful model of learning is the field-tested cognitive apprenticeship model [28]. According to this model, students can learn effectively if the instructional design involves three essential components: “modeling”, “coaching and scaffolding”, and “weaning”. In this approach, “modeling” means that the instructional approaches demonstrate and exemplify the criteria for good performance and the skills that students should learn (e.g., how to solve physics problems systematically). “Coaching and scaffolding” means that students receive appropriate guidance and support as they actively engage in learning the content and skills necessary for good performance. “Weaning” means reducing the support and feedback gradually to help students develop self-reliance [28]. In traditional physics instruction, especially at the college level, there is often a lack of coaching and scaffolding: students come to class where the instructor lectures and does some example problems, then students are left on their own to work through homework with little or no feedback. This is akin to a piano instructor demonstrating for the students how to play the piano and then asking students to go home and practice. This lack of prompt feedback and scaffolding can be detrimental to learning.

Some of the commonly used evidence-based active-engagement (EBAE) approaches implemented in physics include peer instruction with clickers popularized by Eric Mazur from Harvard University [29-32], tutorial-based instruction in introductory and advanced courses [33-35] and collaborative group problem solving [36-39], e.g., using context-rich problems [11-12]. In all of these evidence-based approaches, formative assessment plays a critical role in student learning [40]. Formative assessment tasks are frequent, low-stakes assessment activities which give feedback both to students as well as instructors about what students have learned at a given point. Using frequent formative assessments helps make the learning goals of the course concrete to students, as well as provides them with a way to track their progress in the course with respect to these learning goals. When formative assessment tasks such as concept-tests, tutorials and collaborative group problem solving are interspersed throughout the course, the distinction between teaching and learning is blurred [40-41].

Moreover, technology is increasingly being exploited for pedagogical purposes to improve student learning. For example, Just-in-Time Teaching (JiTT) is an instructional approach in which instructors receive feedback from students before class and use that feedback to tailor in-class instruction [42-44]. Typically, students complete an electronic pre-lecture assignment in which they give feedback to the instructor regarding any difficulties they have had with the assigned reading material, lecture videos, and/or other self-paced instructional tools. The instructor then reviews student feedback before class and makes adjustments to the in-class activities. For example, Eric Mazur's Perusall system [45] allows students to read the textbook and ask questions electronically and the system uses their questions to draft a "confusion report" which distills their questions to three most common difficulties. Then, during class, students may engage in discussions with the instructor and with their classmates, and the instructor may then adjust the next pre-lecture assignment based on the progress made during class. It has been hypothesized that JiTT may help students learn better because out-of-class activities cause students to engage with and reflect on the parts of the instructional material they find challenging. In particular, when the instructor focuses on student difficulties in lecture which were found via electronic feedback before class, it may create a "time for telling" [46] especially because students may be "primed to learn" better when they come to class if they have struggled with the material during pre-lecture activities. The JiTT approach is often used with peer discussion and/or collaborative group problem solving inter-dispersed with lectures in the classroom.

In addition, in the last decade, the JiTT pedagogy has been extended a step further with the maturing of technology [47-66] and "flipped" classes with no in-class lectures have become common with instructors asking students to engage with short lecture videos and concept questions associated with each video outside of the class and using the entire class-time for active-engagement. The effectiveness of flipped classes in enhancing student learning can depend on many factors including the degree to which evidence-based pedagogies that build on students' prior knowledge and actively engage them in the learning process are used, whether there is sufficient buy-in from students and the incentives that are used to get students engaged with the learning tools both inside and, equally importantly, outside the classroom.

Moreover, research suggests that effective use of peer collaboration can enhance student learning in many instructional settings in physics classes including in JiTT and flipped environments, and with various types and levels of student populations. Although the details of implementation vary, students can learn from each other in many different environments. Integration of peer interaction with lectures has been popularized in the physics community by Mazur. In Mazur's approach [67], the instructor poses concrete conceptual problems in the form of conceptual multiple-choice clicker questions to students throughout the lecture and students discuss their responses with their peers. Heller et al. have shown that collaborative problem solving with peers in the context of quantitative "context-rich" problems [11-12] can be valuable both for learning physics and for developing effective problem solving strategies.

Cognitive apprenticeship [28] is one framework that can be used to understand why the EBAE instructional strategies that take advantage of peer discussion and collaboration may be successful in helping students learn. The EBAE pedagogies provide instructors with an opportunity to receive feedback on common student difficulties. The instructors often use this feedback to adjust their in-class activities to effectively build on students' prior knowledge, thus providing students with the necessary coaching and scaffolding to help them learn. Peer discussion also provides students with an opportunity to be coached by their peers who may be able to discern their difficulties even better than the instructor, and carefully designed targeted feedback from the instructor after the peer discussion can provide appropriate scaffolding.

### B. Focus of our research: Comparing introductory physics student performance in EBAE (flipped and non-flipped) courses with LB courses:

In this study, we used the Force Concept Inventory (FCI) [71] in the first semester courses and the Conceptual Survey of Electricity and Magnetism (CSEM) [72] in the second semester courses to assess student learning. The FCI, CSEM and other standardized physics surveys [71-78] have been used to assess introductory student understanding of physics concepts by a variety of educators and physics education researchers [79]. One reason for their extensive use is that many of the items on the survey have strong distractor choices which correspond to students' common difficulties so students are unlikely to answer the survey questions correctly without having good conceptual understanding. In the research discussed here, the performance of students in EBAE courses at a particular level is compared with primarily LB courses in two situations: (I) the same instructor taught two courses, one of which was a flipped course involving EBAE methods and the other an LB course, while the homework and final exams were kept the same, (II) student performance in all of the EBAE courses taught by different instructors were averaged and compared with primarily LB courses of the same type also averaged over different instructors. Whenever differences between these two groups were observed (with students in EBAE courses performing better than students in the LB courses), we investigated which students were benefitting most from the EBAE courses, e.g., those who performed well or poorly on the pretest given at the beginning of the course.

Finally, we were also interested in the typical correlation between the performance of students on the validated conceptual surveys and their performance on the final exam, which typically places a heavy weight on quantitative physics problems.

### C. Framework used in our research exploring the effectiveness of EBAE pedagogies:

We compare introductory physics student performance in EBAE flipped and active-engagement non-flipped courses with LB courses with inspiration from several theoretical frameworks. The overarching framework that is used for the instructional design of all of the EBAE courses in this study (whether flipped or active-engagement non-flipped) was the cognitive apprenticeship model [28,80,81]. This framework focuses on providing opportunities to coach students and scaffold their learning. All of the EBAE classes were designed to give students similar coaching and scaffolding to develop their problem solving and reasoning skills. The EBAE courses focused on the cognitive approach to instructional design for various learning units and building on students' prior knowledge in order to help them learn better. For example, Piaget's framework [82], which emphasizes "optimal mismatch" between what a student knows and where the instruction should be targeted in order for desired assimilation and accommodation of knowledge to occur, was helpful in developing the instructional design. A related framework is the theory of conceptual change put forth by Posner et al. [83]. In this framework, conceptual changes or "accommodations" can occur when the existing conceptual understanding of students is not sufficient for or is inconsistent with new phenomena they are learning about. These frameworks also suggest that these accommodations can be very difficult for students, particularly when students are firmly committed to their prior understanding, unless instructional design explicitly accounts for these difficulties. The model suggests that it is important for instructors to be knowledgeable about student ideas, e.g., which they may apply in inappropriate contexts to make incorrect inferences while solving physics problems. Within this framework, students can be motivated by an anomaly which provides a cognitive conflict and illustrates how their conceptions are inadequate for explaining a newly encountered physical situation, so they become dissatisfied with their current understanding of concepts and improve their understanding. Taking inspiration from these frameworks, EBAE instructors tried to focus on student conceptions and their difficulties in learning physics in order to design instruction that produces the desired cognitive conflict and learning.

## II. METHODOLOGY

### A. Courses and Participants:

The participants in this study were students in 16 different algebra-based and calculus-based introductory physics courses (more than 1500 students in first semester courses and more than 1200 students in the second semester courses) at a typical large research university in the US (University of Pittsburgh). The courses fall into three categories:

1) A lecture-based (or LB) course is one in which the primary mode of instruction was via lecture. In addition to the three or four weekly hours for lectures, students attended an hour long recitation section taught by a graduate TA. In recitation, the TA typically answered student questions (mainly about their homework problems which were mostly textbook style quantitative problems), solved problems on the board and gave students a quiz in the last 10-20 minutes.
2) A flipped course is one in which the class was broken up into two almost equal size groups with each group meeting with the instructor for half the regular class time. For example, for a 200 student class scheduled to meet for four hours each week (on two different days), the instructor met with half the class (100 students) on the first day and the other half on the second day. This was possible in the flipped classes since the total contact hours for each instructor each week with the students was the same as in the corresponding LB courses. Students watched the lecture videos before coming to class and answered some conceptual questions which were based upon the lecture video content. They uploaded the answers to those conceptual questions before class onto the course website and were graded for a small percentage of their grade (typically 4-8%). Although students had to watch several videos outside of class in preparation for each class, each video was typically 5-10 minutes long, followed by concept questions. On average, students in a flipped class had to watch recorded videos which took a little less than half the allotted weekly time for class (e.g., for the courses scheduled for four hours each week, students watched on average 1.5 hours of videos each week, and in the courses scheduled for three hours each week, students watched around one hour of videos). These video times do not include the time that students would take to rewind the video, stop and think about the concepts and answer the concept questions embedded after the videos that counted for their course grade. In the spirit of JiTT, the instructors of the flipped courses adjusted the in-class activities based upon student responses to online concept questions which were supposed to be submitted the night before the class. About 90% of the students submitted their answers to the concept questions that followed the videos to the course website before coming to the class. The web-platforms used for managing, hosting and sharing these videos and for having online discussions with students about them asynchronously (in which students and the instructor participated) were Classroom Salon or Panopto. In-class time was used for clicker questions involving peer discussion and then a whole class discussion of the concept-tests, collaborative group problem solving involving quantitative problems in which 2-3 students worked in a group (followed by a clicker question about the order of magnitude for the answer to the quantitative problem on which students worked collaboratively) and lecture-demonstrations with preceding clicker questions on the same concepts. In addition to the regular class times, students attended an hour long recitation section which was taught the same way as for students in the LB courses.

It is important to note that the instructors who taught the flipped courses also taught LB courses at the same time (usually teaching two courses in a particular semester: one flipped and one LB). Students in both flipped and LB courses completed the same homework and took the same final exam. For the calculus-based flipped courses, the students also took the same midterm exams. This was not possible for the algebra-based courses because the exams were scheduled at different times. However, in the algebra-based courses they took the same final exam and had the same homework. Additionally, the instructors attempted to make the actual delivery of content (done via videos in the flipped courses and via in-class lecture in the lecture-based courses) very similar. Essentially, the content of the videos was delivered in-class in the lecture based courses.

3) EBAE interactive non-flipped course. In this course, the instructor combined lectures with research-based pedagogies including clicker questions with peer discussion, conceptual tutorials, collaborative group problem solving, and lecture demonstrations with preceding clicker questions on the same concepts similar to the flipped courses. In addition, students attended a reformed recitation which primarily used context-rich problems to get students to engage in group problem solving or worked on research-based tutorials while being guided by a TA. The instructor ensured that the problems students solved each week in the recitation activities were closely related to what happened in class. Students also worked on some research-based tutorials during class in small groups, but if they did not finish them in the allotted time, they were asked to complete them as homework.

### B. Materials

The materials used in this study are the conceptual FCI and CSEM multiple-choice (five choices for each question) standardized surveys which were administered in the first week of classes before instruction in relevant concepts (pretest) and the after instruction in relevant concepts (posttest). Apart from the data on these surveys that the researchers collected from all of these courses, each instructor administered his/her own final exam which was mostly quantitative (60%-90% of the questions were quantitative although some instructors had either the entire final exam or part of it in a multiple-choice format with five options for each question to make grading easier). Ten course instructors (who also provided the FCI or CSEM data from their classes) provided their students' final exam scores and most of them also provided a copy of their final exam.

### C. Methods:

Our main goal in this investigation was to compare the average performance of students in introductory physics courses that used EBAE pedagogies with the average performance of students in LB courses by using standardized conceptual surveys, the FCI (for physics I) and CSEM (for physics II) as pre-/posttests. We not only calculated the average gain (posttest - pretest scores) for each group but also calculated the average normalized gain, which is commonly used to determine how much the students learned from pretest to posttest taking into account their initial scores on the pretest. It is defined as $\langle g \rangle = \frac{\%\langle S_f \rangle - \%\langle S_i \rangle}{100 - \%\langle S_i \rangle}$ , in which $\langle S_f \rangle$ and $\langle S_i \rangle$ are the final (post) and initial (pre) class averages, respectively. Then, $Norm\ g = 100\langle g \rangle$ in percent [16]. This normalized gain provides valuable information about how much students have learned by taking into account what they already know based on the pretest. We wanted to investigate whether the normalized gain is higher in one course compared to another.

In order to compare EBAE courses with LB courses, we performed *t*-tests [84] on FCI or CSEM pre- and posttest data. We also calculated the effect size in the form of Cohen's $d$ defined as $(\mu_1 - \mu_2)/\sigma_{pooled}$, where $\mu_1$ and $\mu_2$ are the averages of the two groups being compared (e.g., EBAE vs. LB) and $\sigma_{pooled} = \sqrt{({\sigma_1}^2 + {\sigma_2}^2)/2}$ (here $\sigma_1$ and $\sigma_2$ are the standard deviations of the two groups being compared).

Moreover, although we did not have control over the type of final exam each instructor used in his/her courses, we wanted to look for correlation between the FCI/CSEM posttest performance and the final exam performance for different instructors in the algebra-based and calculus-based EBAE or LB courses. Including both the algebra-based and calculus-based courses, 10 instructors provided the final exam scores for their classes. We used these data to obtain linear regression plots between the posttest and the final exam performance for each instructor and computed the correlation coefficient between the performance of students on the validated conceptual surveys and their performance on the final exam for different instructors. These correlation coefficients between the conceptual surveys and the final exam (with strong focus on quantitative problem solving) can provide an indication of the strength of the correlation between conceptual and quantitative problem solving in introductory physics courses.

Out of all introductory physics courses (algebra-based or calculus-based physics I or II) included in this study, there were four EBAE courses: two completely flipped classes in algebra-based introductory physics I and one completely flipped and one interactive active engagement class in calculus-based introductory physics II.

## III. RESULTS

Table I shows the intra-group pre-/posttest data (pooled data for the same type of courses) on the FCI survey for the calculus-based and algebra-based physics I courses. For the algebra-based courses, some were EBAE courses while others were LB courses, whereas all the calculus-based courses were LB. We find statistically significant improvements from the pretest to the posttest for each group but the normalized gain (Norm g) is largest (30%) for the EBAE courses.

Table I. Intra-group FCI pre-/posttest averages (Mean) and standard deviations (SD) for first-semester introductory physics in calculus-based LB courses, and algebra-based EBAE (flipped) and LB courses. The number of students in each group, N, is shown. For each group, a *p*-value obtained using a *t*-test shows that the difference between the pre-/posttest is statistically significant and the normalized gain (Norm g) from pre- to posttest shows how much students learned from what they did not already know based on the pretest.

| Type of class | FCI | N | Mean | SD | p-value | Norm g |
|---|---|---|---|---|---|---|
| Calc LB | Pretest | 461 | 51% | 21% | <0.001 | 25% |
| | Posttest | 350 | 63% | 20% | | |
| Alg flipped | Pretest | 299 | 35% | 18% | <0.001 | 30% |
| | Posttest | 262 | 54% | 20% | | |
| Alg LB | Pretest | 837 | 35% | 17% | <0.001 | 23% |
| | Posttest | 738 | 50% | 19% | | |

Table II shows the intra-group (pooled data for the same type of courses) pre-/posttest data on the CSEM survey for algebra-based and calculus-based introductory physics II courses. We find that there are statistically significant differences between the pre-/posttest scores for each group but the normalized gain (Norm g) is largest (36%) for the EBAE courses.

Table II. Intra-group CSEM pre-/posttest averages (Mean) and standard deviations (SD) for second- semester introductory physics in calculus-based LB and EBAE courses (here, EBAE flipped and interactive non-flipped courses are combined) and algebra-based LB courses. The total number of students in each group, N, is shown. For each group, a *p*-value obtained using a *t*-test shows that the difference between the pre-/posttest s is statistically significant and the normalized gain (Norm g) from pretest to posttest shows how much students learned from what they did not already know based on the pretest.

| Type of Class | CSEM | N | Mean | SD | p-value | Norm g |
|---|---|---|---|---|---|---|
| Calc LB | Pretest | 410 | 38% | 14% | <0.001 | 21% |
| | Posttest | 346 | 51% | 17% | | |
| Calc EBAE | Pretest | 346 | 37% | 16% | <0.001 | 36% |
| | Posttest | 300 | 60% | 19% | | |
| Alg LB | Pretest | 514 | 24% | 11% | <0.001 | 25% |
| | Posttest | 449 | 43% | 17% | | |

Table III shows the inter-group FCI pre-/posttest score comparison between algebra-based LB and EBAE courses, first holding the instructor fixed (same instructor taught both the LB and EBAE courses, used the same homework and final exams) and second, combining all instructors who used similar methods in the same group (only one instructor used EBAE methods, but several who taught LB courses were combined). Table III shows that there is no statistically significant difference between the pretest scores of students in the LB and EBAE courses in introductory physics 1 on the FCI. Table III also shows that the effect sizes for comparing FCI posttest performance of students in EBAE courses with students in LB courses are 0.314 (same instructor teaching both courses in the same semester) and 0.233 when different courses using similar methods are combined (which are considered small effect sizes).

Table IV shows the inter-group CSEM pre-/posttest score comparison between calculus-based LB and EBAE courses, first holding the instructor fixed (same instructor taught both the LB and EBAE courses and used the same homework and final exams) and second, combining all instructors who taught using similar methods in the same group. Table IV shows that there is no statistically significant difference between the pretest scores of students in the LB and EBAE courses in introductory physics II on the CSEM. Table IV also shows that the effect sizes for comparing CSEM posttest performance of students in EBAE courses with students in LB courses are 0.357 (same instructor teaching both courses) and 0.494 when different courses using similar methods are combined (which are considered medium effect sizes).

Table III. Inter-group comparison of the average FCI pre-/posttest scores of algebra-based students in LB courses with EBAE courses when (i) both courses are taught by the same instructor and (ii) different instructors using similar instructional methods are combined. The p-values and effect sizes are obtained when comparing the LB and EBAE courses in terms of students' FCI scores.

| Comparison of LB and EBAE Groups | | FCI-Pre | FCI-Post |
|---|---|---|---|
| | LB | N: 466<br>Mean: 35%<br>SD: 17% | N: 433<br>Mean: 48%<br>SD: 20% |
| (i) FCI Alg: LB vs. EBAE<br>(same instructor) | Comparison<br>LB and EBAE | p-value: 0.831<br>effect size:<br>0.017 | p-value: <0.001<br>effect size:<br>0.314 |
| | EBAE | N: 262<br>Mean: 35%<br>SD: 18% | N: 262<br>Mean: 54%<br>SD: 20% |
| | LB | N: 837<br>Mean: 35%<br>SD: 17% | N: 738<br>Mean: 50%<br>SD: 19% |
| (ii) FCI Alg: LB vs. EBAE<br>(different instructors combined) | Comparison<br>LB and EBAE | p-value: 0.901<br>effect size:<br>0.009 | p-value: 0.001<br>effect size:<br>0.233 |
| | EBAE | N: 299<br>Mean: 35%<br>SD: 18% | N: 262<br>Mean: 54%<br>SD: 20% |

Table IV. Inter-group comparison of the average CSEM pre-/posttest scores of calculus-based students in LB courses with EBAE courses when (i) both courses are taught by the same instructor and (ii) different instructors using similar instructional methods are combined. The p-values and effect sizes are obtained when comparing the LB and EBAE courses in terms of students' CSEM scores.

| Comparison of LB and EBAE Groups | | CSEM Pre | CSEM Post |
|---|---|---|---|
| | LB | N: 178<br>Mean: 40%<br>SD: 13% | N: 154<br>Mean: 48%<br>SD: 15% |
| (i) CSEM Calc: LB vs. EBAE<br>(same instructor) | Comparison<br>LB and<br>EBAE | p-value:<br>0.895<br>effect size:<br>0.013 | p-value: 0.001<br>effect size: 0.357 |
| | EBAE | N: 208<br>Mean: 40%<br>SD: 15% | N: 181<br>Mean: 54%<br>SD: 19% |
| | LB | N: 410<br>Mean: 38%<br>SD: 14% | N: 346<br>Mean: 51%<br>SD: 17% |
| (ii) CSEM Calc LB vs. EBAE<br>(different instructors pooled) | Comparison<br>LB and<br>EBAE | p-value:<br>0.886<br>effect size:<br>0.011 | p-value: <0.001<br>effect size: 0.494 |
| | EBAE | N: 346<br>Mean: 37%<br>SD: 16% | N: 300<br>Mean: 60%<br>SD: 19% |

Table V shows the average FCI pre-/posttest scores for algebra-based and CSEM pre-/posttest scores for calculus-based courses (Av-Pre/Post), gain (Post-Pre), normalized gain (Norm g), and final exam scores (Av-Fin) for students in the flipped and LB courses taught by the same instructor (with the same homework and final exam) with students divided into three groups based on their pretest scores. A closer look at the gains and normalized gains for the courses taught by the same instructor shows that students in all of the three pretest score categories in the flipped courses had higher gains and normalized gains compared to those in the LB courses taught by the same instructor. Moreover, for algebra-based physics 1, the average final exam scores of the students in the flipped course taught by the same instructor in all the three pretest categories are somewhat higher than the LB course.

Table VI shows the average FCI pre-/posttest scores for algebra-based and calculus-based courses (Av-Pre/Post), gain (Post-Pre), and normalized gain (Norm g) for students in the flipped and LB courses with students divided into three groups based on their pretest scores. All equivalent (algebra-based or calculus-based physics I) courses which used the same instructional strategy (flipped or LB) were combined and students were divided into three groups based upon their pretest scores. A closer look at the gains and normalized gains for the algebra-based courses (for which there are both flipped and LB groups) shows that students in all of the three pretest score categories in the flipped courses had higher gains and normalized gains compared to those in the traditional courses. In the calculus-based LB courses, the highest one-third of the students had 83% and 82% as their FCI pretest and posttest scores, respectively. In Table VI, we do not list the average final exam performance since instructors used different exams which varied in difficulty.

Table V. Average FCI pre-/posttest scores for algebra-based and CSEM pre-/posttest scores for calculus-based courses (Av-Pre/Post), Gain (Post – Pre), normalized gain (Norm g) and final exam scores (Av-Fin) for students in the flipped and LB courses taught by the same instructor (with same homework and final exam) with students divided into three groups based on their pretest scores as shown. Students in the LB or flipped courses in the shaded region can be compared with each other and those in the unshaded region can be compared with each other.

| | Pretest Split | Av-Pre | Av-Post | Gain | Norm g | Av-Fin |
|---|---|---|---|---|---|---|
| FCI Alg LB (Instructor 1) | bottom 1/3 | 18 | 36 | 18 | 22 | 48 |
| | middle 1/3 | 32 | 45 | 13 | 20 | 54 |
| | top 1/3 | 54 | 66 | 12 | 27 | 65 |
| FCI Alg Flipped (Instructor 1) | bottom 1/3 | 17 | 41 | 24 | 29 | 54 |
| | middle 1/3 | 32 | 49 | 17 | 25 | 54 |
| | top 1/3 | 56 | 74 | 18 | 40 | 65 |
| CSEM Calc LB (Instructor 2) | bottom 1/3 | 26 | 35 | 9 | 12 | 43 |
| | middle 1/3 | 39 | 46 | 8 | 12 | 53 |
| | top 1/3 | 53 | 60 | 6 | 14 | 59 |
| CSEM Calc Flipped (Instructor 2) | bottom 1/3 | 25 | 42 | 18 | 24 | 51 |
| | middle 1/3 | 39 | 49 | 11 | 18 | 56 |
| | top 1/3 | 58 | 70 | 12 | 29 | 69 |

Table VII shows the average CSEM pre-/posttest scores for algebra-based and calculus-based courses (Av-Pre/Post), gain (Post-Pre), and normalized gain (Norm g) for students in the EBAE and LB courses with students divided into three groups based on their pretest scores. All equivalent (algebra-based or calculus-based physics II) courses which used the same instructional strategy (EBAE or LB) were combined and students were divided into three groups based upon their pre-test scores. A closer look at the gains and normalized gains for the calculus=based courses (for which there are both EBAE and LB groups) shows that students in all of the three pretest score categories in the EBAE courses had higher gains and normalized gains than those in the traditional courses.

Table VI. Average FCI pre-/posttest scores (Av-Pre/Post), Gain (Post – Pre), and normalized gain (Norm g) for students in the flipped and LB algebra-based and calculus-based courses. All courses in the same group were combined with students divided into three groups based upon their pretest scores as shown. Students in the LB or flipped courses in the shaded region can be compared with each other.

| | Pretest Split | Av-Pre | Av-Post | Gain | Norm g |
|---|---|---|---|---|---|
| FCI Calc LB | bottom 1/3 | 31 | 46 | 15 | 22 |
| | middle 1/3 | 55 | 68 | 13 | 28 |
| | top 1/3 | 83 | 82 | -1 | -7 |
| FCI Alg Flipped | bottom 1/3 | 17 | 41 | 24 | 29 |
| | middle 1/3 | 32 | 49 | 17 | 25 |
| | top 1/3 | 56 | 74 | 18 | 40 |
| FCI Alg LB | bottom 1/3 | 19 | 35 | 16 | 19 |
| | middle 1/3 | 33 | 46 | 14 | 20 |
| | top 1/3 | 55 | 68 | 14 | 30 |

We should note that differences in normalized gain should be interpreted carefully because we do not have a measure of the variability of normalized gain, and thus, differences of 5% may or may not be significant. We stress that we are not making statements about significant differences between EBAE courses and LB courses based on normalized gain, and any statements we have made about significant differences are supported by statistical analyses (e.g., Cohen's d, or comparison of post-test results).

Table VII. Average CSEM pre-/posttest scores (Av-Pre/Post), Gain (Post – Pre), and normalized gain (Norm g) for calculus-based students in the EBAE and LB courses and algebra-based students in LB courses. All courses in the same group were combined with students divided into three groups based upon their pretest scores as shown. Students in the LB or flipped courses in the shaded region can be compared with each other.

| | Pretest Split | Av-Pre | Av-Post | Gain | Norm g |
|---|---|---|---|---|---|
| Calc EBAE CSEM | bottom 1/3 | 22 | 51 | 29 | 37 |
| | middle 1/3 | 35 | 57 | 22 | 34 |
| | top 1/3 | 56 | 70 | 15 | 33 |
| Calc LB CSEM | bottom 1/3 | 23 | 39 | 16 | 20 |
| | middle 1/3 | 35 | 47 | 12 | 19 |
| | top 1/3 | 51 | 59 | 8 | 16 |
| Alg LB CSEM | bottom 1/3 | 15 | 36 | 22 | 25 |
| | middle 1/3 | 23 | 44 | 21 | 27 |
| | top 1/3 | 35 | 51 | 16 | 25 |

Figure 1 shows the CSEM posttest performance along with the final exam performance for three different instructors in flipped and LB calculus-based courses (one instructor taught an EBAE and an LB course, two instructors taught LB courses). Fig. 1 shows that the linear regressions [84] for the flipped and LB courses are fairly similar and that there is a moderate correlation between CSEM posttest scores and final exam scores. We also plotted linear regressions for the algebra-based courses, but the data look similar to Fig. 1 and so are not included here. Instead, we include all the correlation coefficients (CSEM posttest vs. final exam) for all the courses for which we managed to obtain posttest data. Table VIII summarizes the correlation coefficients between post-CSEM/FCI and final exam scores for each instructor who provided final exam data.

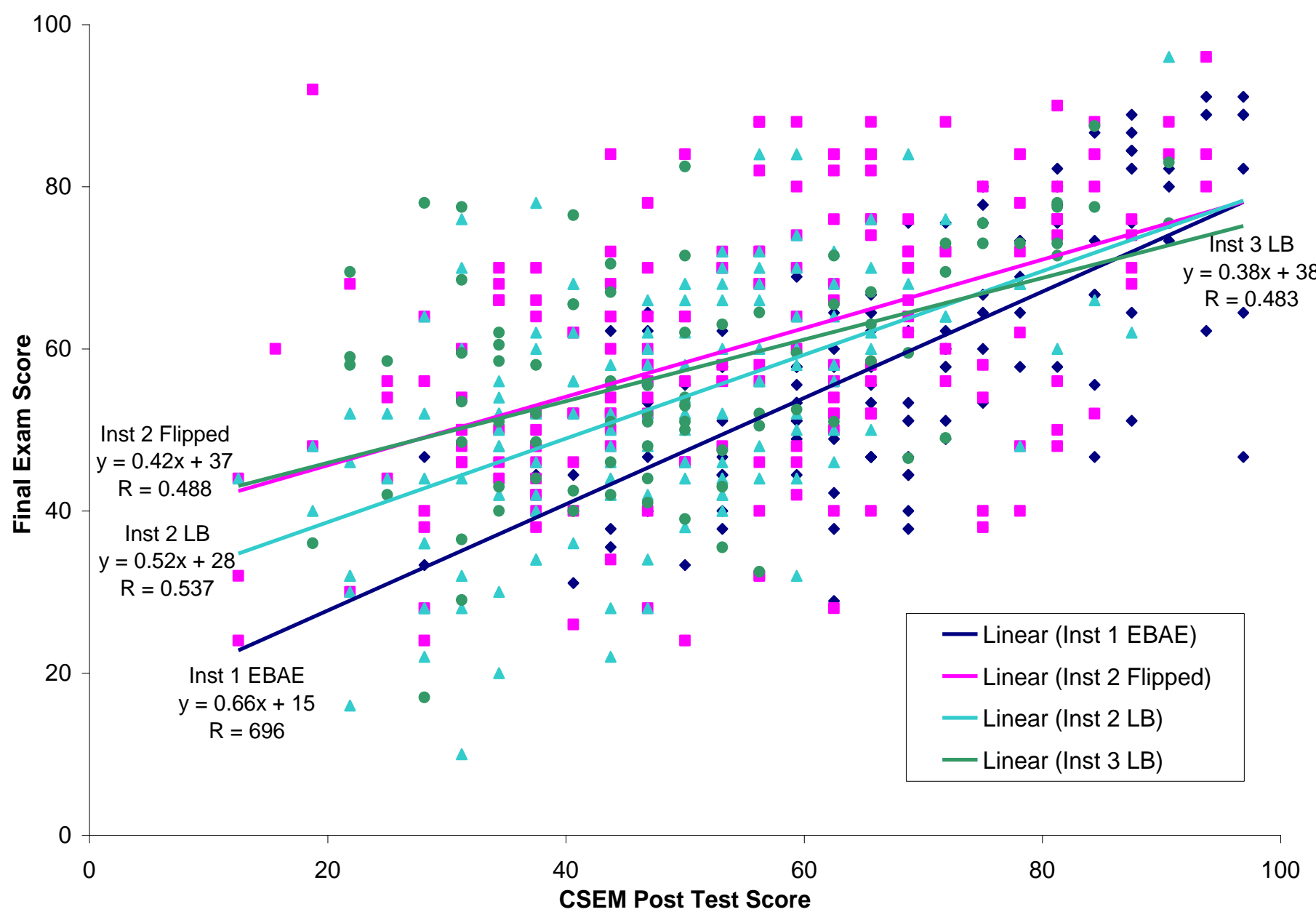

Fig 1. Linear regression of the CSEM posttest scores (conceptual) and final exam scores (heavy focus on quantitative problems) for four calculus-based introductory physics courses shows the correlation coefficients between 0.438-0.598. There were no clear trends in the correlation coefficients based upon whether the instructor (Inst) used EBAE strategies or whether the class was LB.

Table VIII: Correlation coefficients (R) between post-CSEM/FCI and final exam scores for each instructor (Inst) who provided final exam data. The final exam data were not provided by physics 2 instructors in algebra-based courses.

| **Physics 1 (Calc)** | | **Physics 1 (Alg)** | | **Physics 2 (Calc)** | |
|---|---|---|---|---|---|
| Instructor & course type | R | Instructor & course type | R | Instructor & course type | R |
| Inst 1: LB | 0.495 | Inst 1: Flipped | 0.559 | Inst 1: EBAE | 0.696 |
| Inst 2: LB | 0.589 | Inst 1: LB | 0.516 | Inst 2: Flipped | 0.488 |
| Inst 3: LB | 0.787 | Inst 2: LB | 0.693 | Inst 2: LB | 0.537 |
| | | | | Inst 3: LB | 0.483 |

## IV. DISCUSSION AND SUMMARY

In all cases investigated, we find that on average, introductory physics students in the courses which made significant use of evidence-based active engagement (EBAE) methods outperformed students in courses primarily taught using lecture-based (LB) instruction on standardized conceptual surveys (FCI or CSEM) in the posttest even though there was no statistically significant difference on the pretest. This was true both in the algebra-based and calculus-based physics 1 (primarily mechanics) and 2 (primarily E&M) courses. Also, the differences between EBAE and LB courses were observed both among students who performed well on the FCI/CSEM pretest (given in the first week of classes) and also those who performed poorly, thus indicating that EBAE instructional strategies help students at all levels.

On the other hand, the typical effect size for the differences between equivalent EBAE and traditional courses is between 0.23-0.49, which is considered small to medium. Thus, the benefits of these EBAE approaches were not as large as one may expect to observe. Why might that be the case and how can instructors enhance student learning more than that observed in this investigation using EBAE instructional strategies?

There are many potential challenges to using EBAE instructional strategies. Below, we list some of the possible challenges and some strategies that may reduce those challenges. Many of these have been described elsewhere [68, 85-97] so we provide only a short summary:

- Lack of student engagement even with well-designed learning tools, which may occur for many different reasons: lack of student motivation, poor self-efficacy or poor time-management skills on the part of the students, lack of effective incentives for students to engage with the self-paced learning tools, etc. Strategies to address some of these difficulties have been described, e.g., providing students with effective strategies to learn [98, 99], using certain communication activities to foster student motivation [85,86]. Other strategies to address these potential issues have also been described [88-90,92-94].

- Lack of student engagement with in-class active learning activities (e.g., group problem solving). Many strategies to help address this issue have been described, e.g., designing in-class activities which foster both individual accountability and positive inter-dependence [11-12]. One example of fostering individual accountability is to include a short quiz or clicker questions related to content students were supposed to learn when working in groups, and positive interdependence means that the success of each student in a group is dependent on the success of others. For other strategies to help foster student engagement see Refs. [85,95,96] and references therein.
- Student misconceptions about learning, or resistance to EBAE instructional strategies which could be addressed at the beginning of the term by framing the instructional design of the class [68,97] and providing data on the effectiveness of the evidence-based strategies being used (and conversely the ineffectiveness of e.g., instructor explanations [100]).
- Large class sizes can be an impediment, and one approach faculty have used in flipped courses is to split the class in two, thus forming smaller class sizes as well as more room for students to form groups and move around the classroom. Undergraduate or graduate teaching assistants can also help in facilitating in-class activities. In group activities, students often work at different rates, and students who finish early can help others.
- Content coverage. There is often a lot of content covered in introductory physics courses and it may be challenging to cover the same amount of content while also including frequent active learning activities during class. Moving some of the content delivery outside of class (e.g., some pre-lecture reading or videos on certain 'easier' concepts, or moving the entire content delivery outside of class like in a flipped course) can help provide additional time for in-class activities.

We note that the instructors who taught the EBAE courses has control of designing the courses themselves and the researchers only provided them with guidance before and during they worked on designing the courses. The instructors may not have addressed some of the potential issues mentioned above sufficiently, e.g., by framing the courses at the beginning of the term, and providing incentives for students to engage both with in-class and out-of-class activities. However, these issues are challenging to fully address, especially in large classes such as those involved in this investigation (as suggested by the data), and iterative refinement of a course is needed in order to address them. Lastly, while we provided the instructors with information about active learning materials developed by physics education researchers, discussions indicated that they adapted or created some of their own materials to fit the way the preferred to teach, and the extent to which the materials they adapted or created are conducive to effective learning is unclear.

In addition, Henderson and Dancy [101] found that many instructors try certain EBAE instructional strategies, but some discontinue use after one or two semesters. The faculty members who persist are usually the ones who get support from their peers (e.g., developing faculty learning communities, working with instructional designers at local teaching and learning centers) because there may be many implementation difficulties specific to a particular university even if a particular EBAE approach has been found to be effective elsewhere. Interacting with others, even from different departments, who have been engaged in evidence-based teaching (e.g., visiting their classes, getting feedback from them about one's own classes, etc.) can be extremely valuable. Often, teaching and learning centers are happy to send someone to observe a class and provide feedback as well as suggestions for future active learning activities.

Furthermore, we note that in this study, we found that student performance in a non-flipped EBAE course (which used active learning interspersed with short lectures in class) was comparable to student performance in a flipped course. It is important to point out that the instructor in this EBAE course ensured that the recitations are effectively used to promote active learning and that the activities used in the recitation were closely tied to the course learning goals. Flipping a course can be a time consuming process especially if the instructor is developing his/her own lecture videos for the first time and he/she has not implemented the EBAE strategies in his/her class earlier. Therefore, it is encouraging to observe that one does not need to flip his/her course completely, but can introduce EBAE activities in regular class and also in recitation. These active learning activities and materials can be modified and improved after each use by getting feedback from the students (and also getting feedback from the TAs teaching recitation).

As discussed earlier, learning gains in EBAE courses were not as high as one might expect. This should not be taken as discouragement, but rather as an indication that effective teaching is an iterative pursuit and one should learn from each course implementation and try to improve. The expectation that introducing a lot of EBAE instructional strategies which have been found to be effective elsewhere will result in large gains without refining the material and implementation can deter instructors from continuing the use of EBAE instructional strategies when the results are less than expected, especially given the time commitment reformed teaching can take initially. Instead, one should continue to make refinement and remember that any improvement in student learning is worth the effort!

In summary, in order to enhance student learning in EBAE classes it important not only to develop effective EBAE learning tools and pedagogies commensurate with students' prior knowledge but also to investigate how to implement them appropriately and how to motivate and incentivize their usage to get buy-in from students in order for them to engage with them as intended. Furthermore, for flipped classes, it is especially important to investigate strategies for having a diverse group of students engage with self-paced learning tools effectively. Investigation of various factors that can deter or incentivize their use is essential in order to develop a holistic learning environment to help students with diverse backgrounds benefit from the self-paced learning tools. Additionally, it will be valuable to examine and compare the effectiveness of self-paced learning tools, e.g., videos and concept questions provided to students in flipped classes, when implemented in a controlled environment in which students must effectively engage with the tool one-on-one in front of a researcher vs. an environment in which students are free to use the tool in whatever manner they choose. A framework for understanding and optimizing the factors that can support or hinder effective use of self-paced learning tools, e.g., those students are asked to engage with in flipped courses, would be helpful in developing and implementing self-paced tools conducive to learning.

## V. ACKNOWLEDGEMENTS

We are grateful to all the faculty and students who helped with the study. We thank the US National Science Foundation for award DUE-1524575.